\documentclass[prb,unsortedaddress,preprint,nobibnotes]{revtex4}

\usepackage{graphics}
\usepackage{amsmath}
\usepackage{subfigure}

\begin{document}

\title{Vacancy clustering and diffusion in silicon: Kinetic lattice Monte Carlo simulations}
\date{\today}

\author{Benjamin P. Haley}
\affiliation{Department of Applied Science, University of California, 
   Davis CA 95616}
\author{Keith M.\ Beardmore}
\affiliation{MZA Associates, Albuquerque NM 87106}
\author{Niels Gr{\o}nbech-Jensen}
\affiliation{Department of Applied Science, University of California, 
   Davis CA 95616}

\begin{abstract}
Diffusion and
clustering of lattice vacancies in silicon as a function of temperature, concentration,
and interaction range are investigated by Kinetic Lattice Monte Carlo simulations.
It is found that higher temperatures lead to larger clusters with
shorter lifetimes on average, which grow by attracting free vacancies, while
clusters at lower temperatures grow by aggregation of smaller clusters.  
Long interaction ranges produce enhanced diffusivity and fewer clusters.
Greater vacancy
concentrations lead to more clusters, with fewer free vacancies, but the size of the
clusters is largely independent of concentration.  
Vacancy diffusivity is shown to obey power
law behavior over time, and the exponent of this law is shown to increase with 
concentration, at fixed temperature, and decrease with temperature, at fixed 
concentration.  
\end{abstract}

\maketitle

\section{Introduction\label{sec_intro}}
The introduction of dopants into crystalline silicon via ion implantation creates damage 
to the crystal lattice.  Lower energy implants, in the 10 - 100 keV range, create lattice 
vacancies and interstitial defects which mostly recombine during annealing so that only 
the implanted ions remain as defects in the lattice\cite{ref_VI_RECOMBINE}.  High energy 
ion implantation, in the MeV range, produces greater physical distance between regions of 
vacancies and interstitials, so that less recombination occurs.  Thus a vacancy layer forms
near the surface, and an interstitial layer forms deeper in the silicon.  The enhanced
diffusion\cite{ref_V_LAYER,ref_V_BINDING} of other defects, such as Sb, after high energy
Si implants, demonstrates the existence of the vacancy layer near the surface.
\par
Vacancies near the surface of silicon can be used to control the diffusion of 
subsequently implanted dopants, such as boron, which are sensitive to transient 
enhanced diffusion. Excess interstitials created by boron implants
recombine with vacancies created by previous high energy Si implants, limiting diffusion 
of the boron dopants\cite{ref_V_B}.  The experimental 
observations\cite{ref_V_LAYER,ref_V_BINDING} of the vacancy layer persist after 
annealing, indicating formation of large stable clusters of vacancies.  Exploring the 
mechanisms by which these voids form is important for understanding the implantation and 
diffusion of dopants in silicon after, e.g., high energy Si implants.
\par
Several theoretical descriptions of the void formation process in silicon exist.  General 
models of Ostwald ripening\cite{ref_OSTWALD}, the growth of large clusters at the expense
of smaller clusters, simulate the dissociation of smaller clusters and the absorption of 
their constituents into larger clusters\cite{ref_MADRAS_MCCOY}.  
Bongiorno and Columbo\cite{ref_BONGIORNO_COLUMBO} used the Stillinger-Weber\cite{ref_SW} 
potential and molecular dynamics simulations to demonstrate that strain fields due to the 
distortion in the silicon lattice near vacancies affect the capture radius in different 
directions from a vacancy cluster.  Staab \textit{et al.}\cite{ref_V_AB_INITIO} performed 
\textit{ab initio} calculations on clusters containing up to 17 vacancies and found that 
those clusters which form closed rings, for example, those with 6 and 10 vacancies, are 
exceptionally stable.  La Magna \textit{et al.}\cite{ref_LA_MAGNA} used lattice
Monte Carlo simulations to study the formation of vacancy clusters.  They considered 
various interactions, including an Ising-like binding model in which only nearest neighbor 
interactions were considered, at $900\ ^{\circ}$C and a high vacancy concentration of 
$10^{20}\ \text{cm}^{-3}$.  They found that a large number of very stable clusters of 6, 
10, and 14 vacancies formed.  They also used an extended Ising-like model which accounted 
for second nearest neighbor interactions.  This extended binding model did not produce 
large numbers of the stable clusters of the nearest neighbor model; the extended model 
resulted in fewer, larger clusters and small clusters were observed to move through the
lattice to form larger clusters, in a self-organizing manner.  Chakravarthi and 
Dunham\cite{ref_CHAK} used a rate equation model to study vacancy cluster growth in silicon
during a 10 minute anneal.  They also observed a high incidence of the stable ring clusters
with 6, 10, and 14 vacancies.  They found that at lower temperature, $750\ ^{\circ}$C, 
smaller clusters with fewer than 36 vacancies were most prominent, while at higher 
temperature, $950\ ^{\circ}$C, larger clusters dominated.  Prasad and Sinno used molecular 
dynamics\cite{ref_PRASAD_SINNO} to model the energetics of vacancy cluster formation and 
used this information to develop a mean-field continuum model\cite{ref_PRASAD_SINNO_2}
of cluster aggregation which also demonstrated the importance of small cluster diffusion in
void formation.
\par
Continuum models such as those of Chakravarthi and Dunham\cite{ref_CHAK} and Prasad and 
Sinno\cite{ref_PRASAD_SINNO,ref_PRASAD_SINNO_2} must assume the behavior of small vacancy 
clusters in silicon with respect to how they aggregate, i.e., whether they diffuse or 
dissociate.  Variation in behavior over ranges of temperature or concentration is not 
considered in these models.  Kinetic Lattice Monte Carlo (KLMC) models do not make 
any assumptions regarding the behavior of clusters since they are based on interactions
between individual defects.  These models can also simulate much 
larger systems and longer times than molecular dynamics simulations, 
and can include arbitrarily long range interactions.  
This work investigates, through a KLMC model, the formation of vacancy 
clusters and the mechanisms by which clusters grow, as a function of temperature, in 
section \ref{sec_V_T}, concentration, in section \ref{sec_V_C}, interaction range, in
section \ref{sec_V_r}.  The time dependence of diffusivity, $D$, as a power law of the form,
\begin{equation}\label{eq_Dt_power} D(t) \sim t^{-\gamma} \end{equation}
is studied in each section.

\section{Simulation method}
\subsection{Kinetic Lattice Monte Carlo}
A Kinetic Lattice Monte Carlo method is used to simulate atomic scale diffusion
processes.  Lattice defects are probed, in random order,
and, if a move to a neighbor site is allowed, attempts to move each defect, one 
at a time.  Defects can include dopants and native point defects, but are
limited to vacancies in this work.  Possible movements therefore include vacancies 
exchanging lattice sites with lattice atoms or other vacancies.  We use the 
Metropolis\cite{ref_MC} algorithm with detailed balance to determine whether a new
configuration is accepted.
If the move does not lower the system energy, it is accepted, with probability 
$e^{-\Delta E/2k_{B}T}$, where $k_B$ is Boltzmann's constant, $T$ is the system 
temperature, and $\Delta E$ is the change in system energy as a result of the move.
The system energy in each configuration is the sum of pair interactions between the 
defects.  The values for the pair interactions at given 
separations on the lattice, out to 18 neighbors,  were calculated with \textit{ab initio}
methods described in section \ref{sec_abinit}.
\par
The time scale of a KLMC simulation is set by the hopping frequency, $\nu_H^V$, of the 
vacancies, which is determined by 
\begin{equation}\label{eq_hopfreq}\nu_H^V = \nu_0^Ve^{-E_b^V/k_BT}\end{equation}
where $\nu_0^V$ is the attempt frequency, $E_b^V$ is the height of the energy barrier for
movement to a nearest neighbor position. In this work, we used $\nu_0^V$ calculated with 
the expression
\begin{equation}\label{eq_attfreq}
\nu_0^V = \frac{8\text{D}_0^V}{\text{a}_0^2},
\end{equation}
and the values $D_0^V = 1.18\times10^{-4}\ \text{cm}^2/\text{s}$,
and $E_b^V = 0.1\ \text{eV}$, reported for low density vacancy diffusion by 
Tang \textit{et al.}\cite{ref_TANG}, and a lattice constant of 5.43 {\AA}.  
This value for $E_b^V$ is less than the 
0.2 eV used by other authors\cite{ref_BUNEA_PRB,ref_BD_MRS}, but the energy barrier for 
the diffusion of a single vacancy should be calculated from a known low density system.
\par
After each KLMC step, which randomly visits defects in the system, the simulation 
time is incremented by the constant time step, which is 
the exchange time of a vacancy with a Si lattice atom, set by $\nu_H^V$, given by
equation \ref{eq_hopfreq}.
Updating the system time allows the following definition of a time dependent diffusion 
coefficient
\begin{equation}\label{eq_Dt}
D^V(t_1,t_2) = \frac{\left\langle\left|r_i(t_2) - r_i(t_1)\right|^2\right\rangle_i}
{6\left|t_2 - t_1\right|}.
\end{equation}
This definition approaches the thermodynamic diffusion coefficient as 
$\left|t_2 - t_1\right|$ approaches infinity,
while also allowing for diffusion studies during various time ranges.

\subsection{\textit{Ab initio} Calculations\label{sec_abinit}}
\begin{figure}\scalebox{0.5}[0.5]{\rotatebox{270}{
\includegraphics{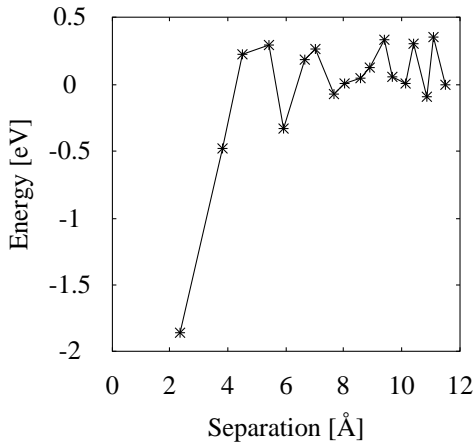}}}
\caption{\label{fig_potentials}
Vacancy pair interaction energies.}
\end{figure}

The pair interactions between vacancies were calculated using the 
ultrasoft pseudopotential plane wave code VASP\cite{ref_VASP}, 
on a 216 atom supercell.  The generalized gradient approximation (GGA) was used, along with 
a $4^3$ Monkhorst-Pack\cite{ref_MPk} \textbf{k}-point sampling and a kinetic energy cutoff 
of 208 eV.  All systems were charge neutral.
The resulting interaction energies are shown in Figure \ref{fig_potentials}. 
In each case a pair of vacancies was placed into the silicon supercell and the initial 
configuration of atoms was allowed to relax.  The final energy of the relaxed configuration
was recorded and the process was repeated for all possible shell separations of the 
vacancy pair out to the cutoff range of 18 shells, the largest separation which could be
practically calculated.  The final energy at the cutoff range 
was shifted to zero, so that the interaction vanishes at the cutoff range,
and all the other values were shifted by the same amount.
These shifted values were recorded as the interaction energies of the vacancy pair.
The process of truncating and shifting the energies results in interactions which seem 
strongly attractive at first and second nearest neighbor positions and weakly repulsive 
at most larger separations.  Thermal fluctuations at elevated temperatures are more likely
to overcome the interactions at long range than at short range, making the dissolution of
clusters less probable than formation.
\par
The set of interaction
energies was incorporated into the KLMC model.  For every configuration of vacancies
in the KLMC simulations the system energy was calculated as the sum of all the pair 
interactions
between the vacancies in the system, given the shell separation of each vacancy pair, out 
to the cutoff interaction range.  Separate vacancies were not permitted to simultaneously 
occupy the same lattice site.

\section{Results}
\begin{figure}[b]
\begin{center}
\subfigure{\scalebox{0.3}[0.3]{\rotatebox{270}{
\includegraphics{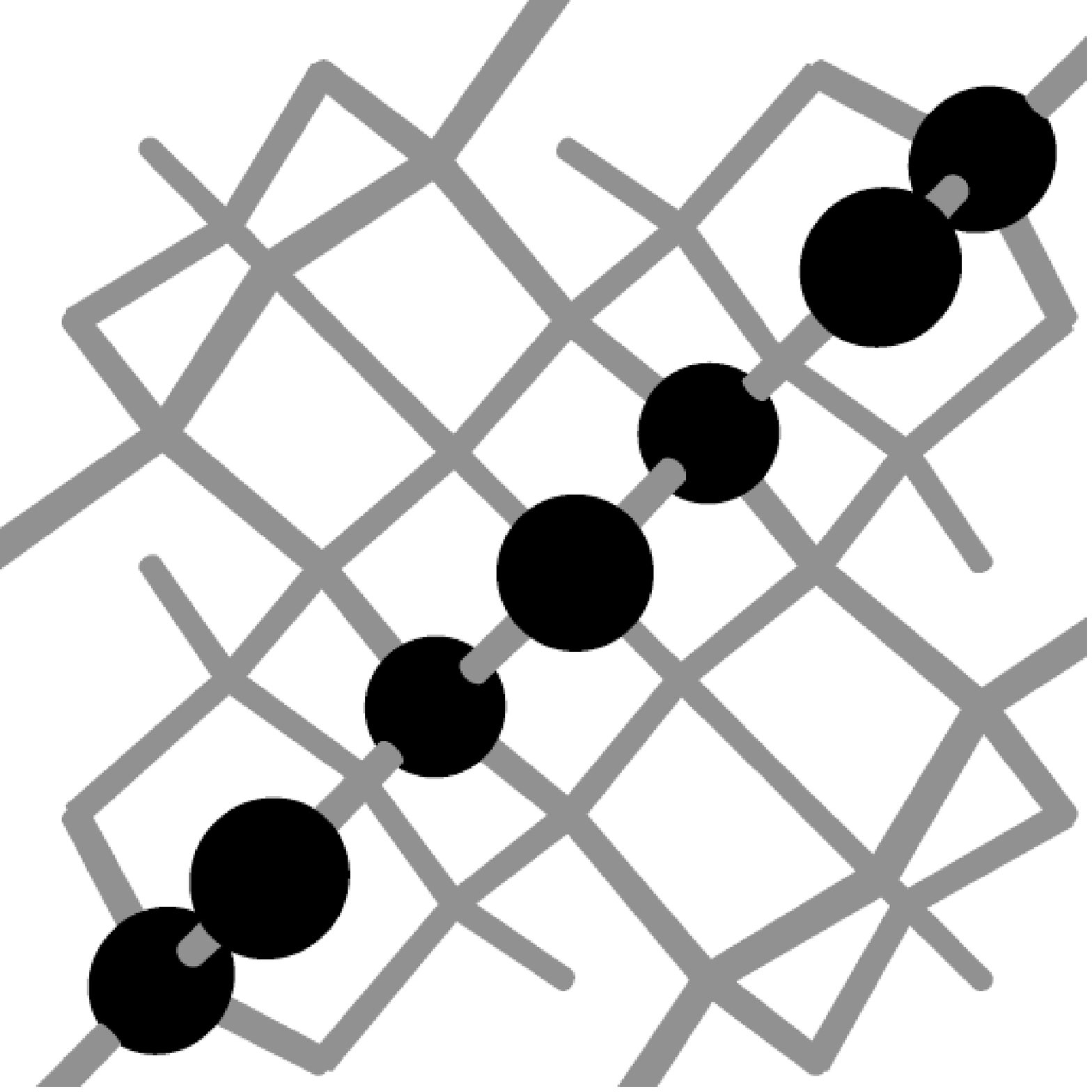}}}}
\hspace{1cm}
\subfigure{\scalebox{0.3}[0.3]{\rotatebox{270}{
\includegraphics{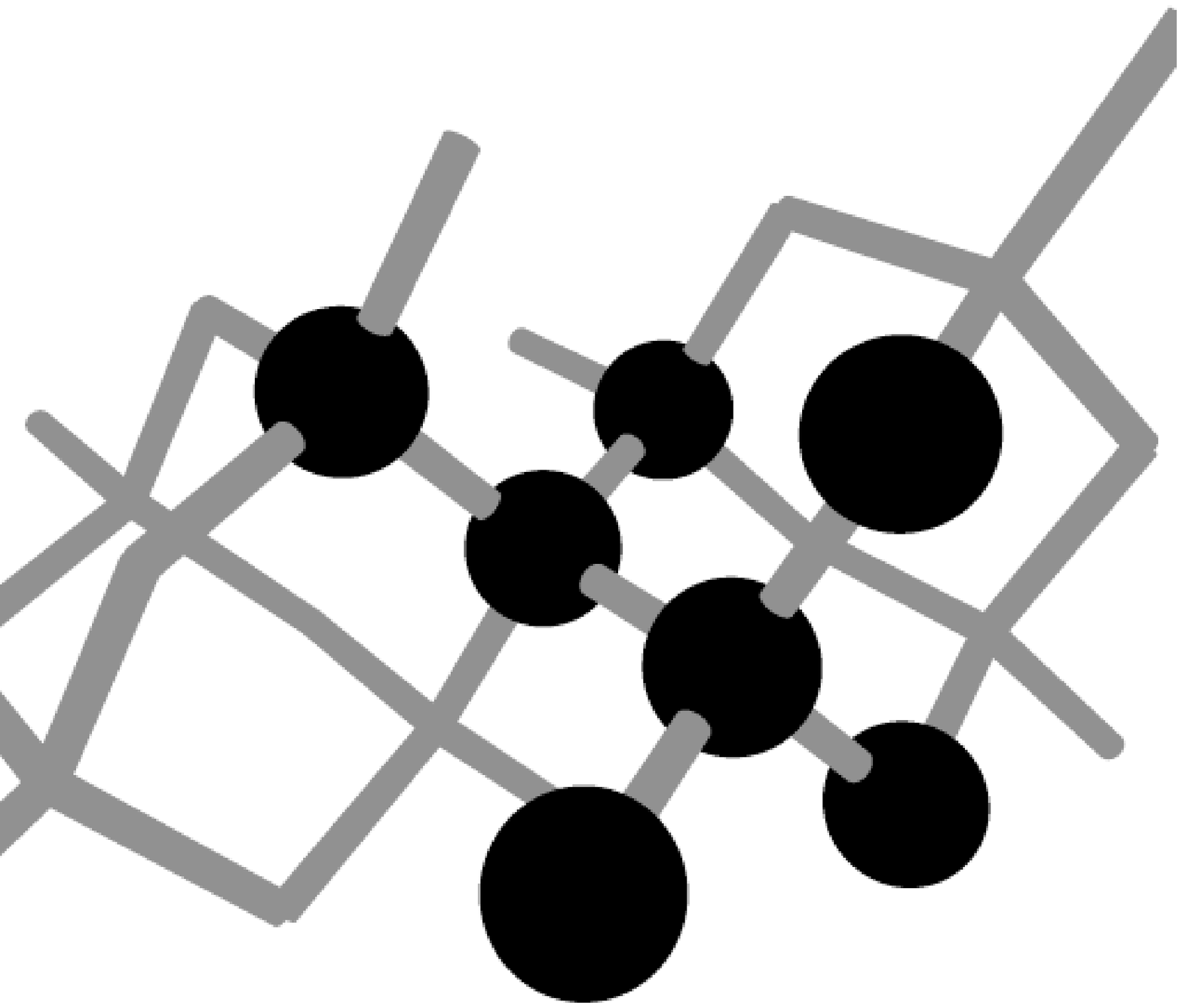}}}}
\caption{\label{fig_clusters}
Vacancy clusters on a silicon lattice in a KLMC simulation. Silicon atoms are not shown.}
\end{center}
\end{figure}

For each combination of
system parameters, five simulations with different random initial distributions of 
vacancies were performed; the mean results are presented here.  Defects were visited 
randomly at each Monte Carlo step.  A defect is a member of a cluster if it is at a
nearest neighbor position to at least one other member of the cluster, as in 
Figure \ref{fig_clusters}.
The boundary conditions of all simulations were periodic.  On all following plots the 
statistical error is smaller than the symbols used, so error bars are not shown.



\subsection{Temperature\label{sec_V_T}}
\begin{figure}\begin{center}\scalebox{0.5}[0.5]{\rotatebox{270}{
\includegraphics{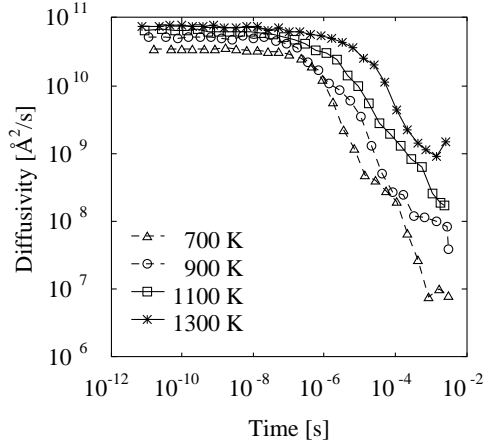}}}
\caption{\label{fig_V_DV_T}
Vacancy diffusivity at various temperatures, with $10^{18}\ \text{V cm}^{-3}$ and 18 shell 
interactions.}
\end{center}\end{figure}

The simulations presented in this section were performed with 170 vacancies in a simulation
box 102 unit cells in each dimension, for a vacancy
concentration of $10^{18}\ \text{cm}^{-3}$.  Long range interactions were used, to the 
eighteenth nearest neighbor, and the temperature was varied from 700 K to 1300 K.
Figure \ref{fig_V_DV_T} shows the diffusivity of vacancies.
The period of initial constant diffusivity, indicating free vacancy 
diffusion, leads to decreasing diffusivity, as a power law with exponent 
$\gamma = 0.8\pm0.2$, as clusters form.
The diffusivity remains greater at higher temperatures, and the power law exponent is 
greater at lower temperatures.
\begin{figure}[!b]\begin{center}\scalebox{0.5}[0.5]{\rotatebox{270}{
\includegraphics{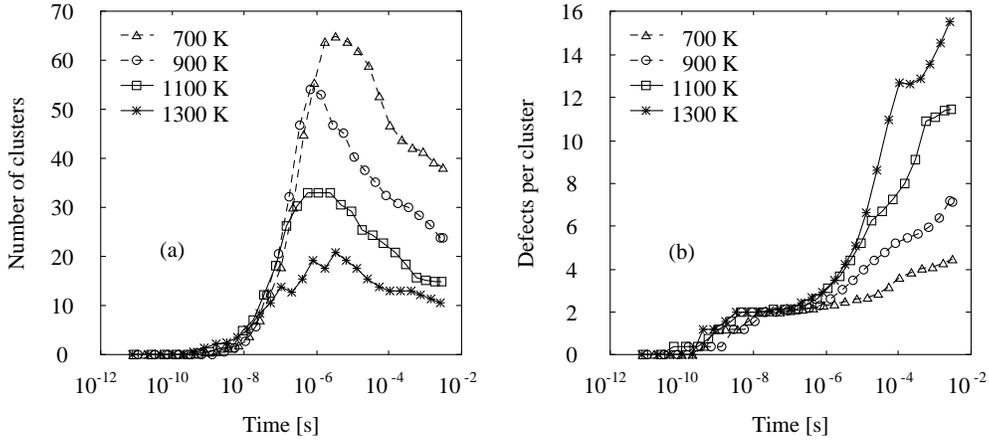}}}
\caption{\label{fig_V_nclust_ndefs_T}
Number of clusters (a) and cluster size (b) at various temperatures, 
with $10^{18}\ \text{V cm}^{-3}$ and 18 shell interactions.}
\end{center}\end{figure}
The formation of vacancy clusters over time can be seen in 
Figure \ref{fig_V_nclust_ndefs_T}a.  At higher temperatures, fewer clusters form but the 
number of clusters varies less over time than at lower temperatures.  The clusters 
which form at higher temperatures tend to be larger on average than those formed at 
lower temperatures, as seen in Figure \ref{fig_V_nclust_ndefs_T}b.  
\begin{figure}[t]\begin{center}\scalebox{0.5}[0.5]{\rotatebox{270}{
\includegraphics{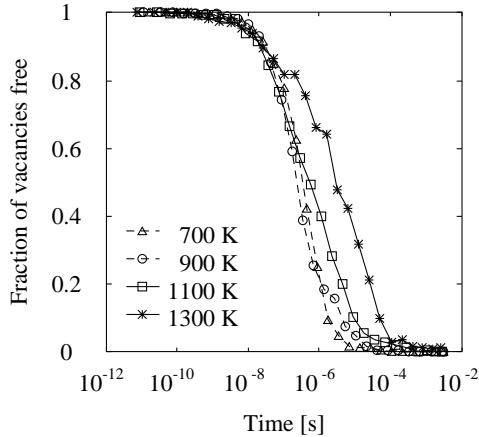}}}
\caption{\label{fig_V_freeV_T}
Fraction of vacancies free at various temperatures, with $10^{18}\ \text{V cm}^{-3}$ and 
18 shell interactions.}
\end{center}\end{figure}
The change in the number of clusters alone can not account for the greater size of the 
clusters formed at higher temperatures.  Figure \ref{fig_V_freeV_T} shows the fraction of 
free vacancies over time for the various temperatures under consideration.  During the time
interval in which the number of clusters is decreasing and the mean cluster size is 
growing, in Figure \ref{fig_V_nclust_ndefs_T}, more vacancies are likely to be free, not 
bound in a cluster, at higher temperatures.  Thus, while the growth of larger clusters at 
all temperatures is partly due to the aggregation of smaller clusters and partly due to
the capture of free vacancies by existing clusters, the latter effect contributes more at 
higher temperatures.  
\par
\begin{figure}\begin{center}\scalebox{0.5}[0.5]{\rotatebox{270}{
\includegraphics{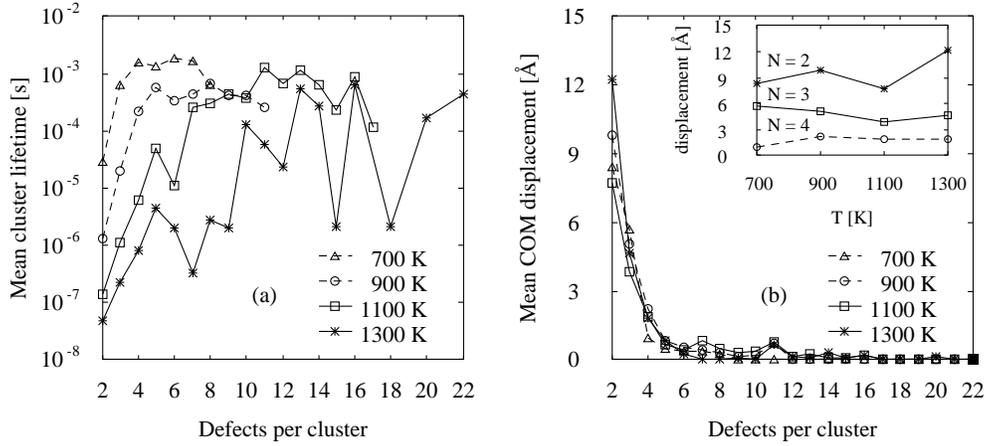}}}
\caption{\label{fig_V_lifetimes_com_T}
Mean cluster lifetime (a) and mean displacement of cluster center (b) at various 
temperatures, with $10^{18}\ \text{V cm}^{-3}$ and 18 shell interactions.}
\end{center}\end{figure}

The mean lifetime of the clusters, defined for each cluster size as the time over which a 
cluster consists of that many defects, also depends on temperature.  
Figure \ref{fig_V_lifetimes_com_T}a shows that the lifetime of clusters at lower 
temperatures is significantly greater than clusters of the same size at higher 
temperatures.  Thermal fluctuations $k_BT$ are larger relative to the vacancy-vacancy 
binding energy at higher temperatures, increasing the probability that a vacancy will 
break free from a cluster.  Certain cluster sizes have longer mean lifetimes than those 
with a similar number of vacancies.  Clusters with 5, 10, 13, and 16 vacancies exist longer
on average than clusters with one more or fewer vacancies, especially at higher 
temperatures.  This trend is similar to the \textit{ab initio} results discussed in 
section \ref{sec_intro}, which noted stable clusters with 6, 10, and 14 vacancies.  It 
should be noted that the \textit{ab initio} results considered many-body effects while the 
KLMC results shown here used only pair interactions.
Figure \ref{fig_V_lifetimes_com_T}b shows the mean displacement of the cluster center for 
all cluster sizes observed in the simulations.  As expected, the smaller clusters are more 
mobile on average, but the inset plot shows that the mean displacement of the three 
smallest clusters, with 2, 3, and 4 vacancies, does not change considerably over the range 
of temperatures considered.  
\par
From the results in 
Figures \ref{fig_V_nclust_ndefs_T}, \ref{fig_V_freeV_T} and \ref{fig_V_lifetimes_com_T},
the growth of large clusters proceeds by different mechanisms at lower temperatures than at
higher temperatures.  At lower temperatures, many small clusters, mostly vacancy pairs, 
form, with most of the vacancies in the system bound in clusters.  The pairs remain bound 
for relatively long times, during which they aggregate into larger clusters.  The rate of 
aggregation is relatively slow because the diffusion of vacancy pairs is slower than the 
diffusion of free vacancies.  At higher temperatures, some vacancy pairs form initially, 
but many vacancies are still free.  The pairs are more likely to dissociate, with the 
constituent vacancies reforming new pairs.  Thus, at higher temperatures the formation of 
large clusters is a combination of aggregating small clusters and capturing 
free vacancies.  The free vacancies diffuse rapidly so larger clusters can form more 
quickly at higher 
temperatures.  The continuum models of void growth discussed in section \ref{sec_intro} 
do not account for this temperature dependence.  The difference in vacancy clustering 
behavior at lower and higher temperatures can be important if the vacancies are used to 
control diffusion of other species.

\subsection{Concentration\label{sec_V_C}}
\begin{figure}\begin{center}\scalebox{0.5}[0.5]{\rotatebox{270}{
\includegraphics{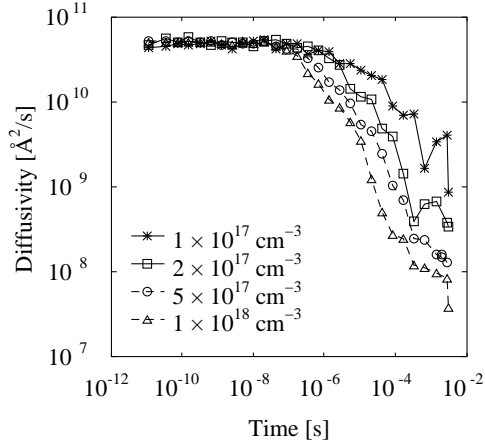}}}
\caption{\label{fig_V_DV_C}
Vacancy diffusivity at various concentrations, with 18 shell interactions at 900K.}
\end{center}\end{figure}

\begin{figure}\begin{center}\scalebox{0.5}[0.5]{\rotatebox{270}{
\includegraphics{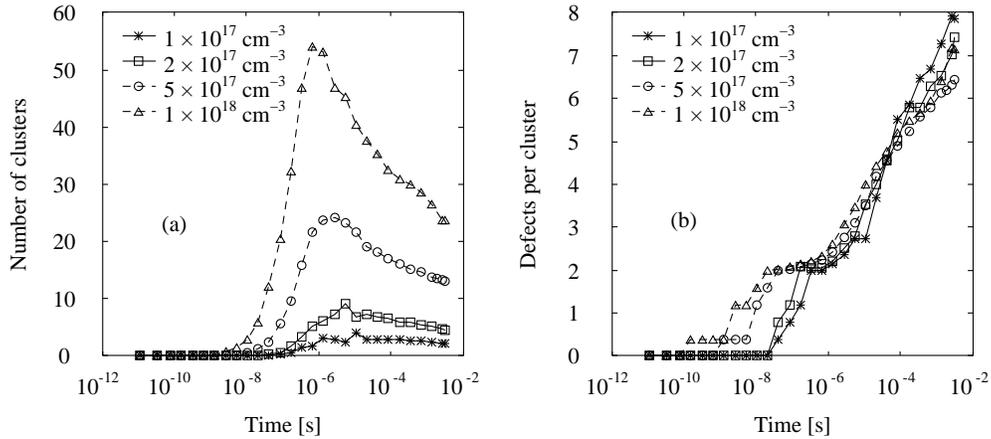}}}
\caption{\label{fig_V_nclust_ndefs_C}
Number of clusters (a) and cluster size (b) at various concentrations, 
with 18 shell interactions at 900 K.}
\end{center}\end{figure}

The simulations presented in this section were performed with a varying number of vacancies 
in a simulation box 102 unit cells in each dimension, for vacancy concentrations ranging 
from $10^{17}\ \text{cm}^{-3}$ to $10^{18}\ \text{cm}^{-3}$.  
Long range interactions were used, to the eighteenth nearest neighbor, and the 
system temperature was 900 K.  Figure \ref{fig_V_DV_C} shows the vacancy diffusivity 
over time.  For all concentrations considered the vacancies 
initially diffuse freely.  The diffusivity decreases as clusters form, with a power law
exponent $\gamma = 0.6\pm0.1$, where the higher values correspond to higher concentration.
The formation of clusters can be seen in Figure \ref{fig_V_nclust_ndefs_C}a.  At lower 
concentrations the number of clusters increases slowly over time, while at higher 
concentrations a large number of clusters form initially, after which the number of 
clusters decreases again.  Figure \ref{fig_V_nclust_ndefs_C}b shows that the clusters, 
which form initially at all concentrations, 
tend to be vacancy pairs and that the mean size of the
clusters increases at roughly the same rate, with slope about 1.4, for all concentrations 
considered.  The 
difference in cluster growth between the lower and higher concentration regimes can be 
understood by considering the number of free vacancies in the system.  
\begin{figure}[!b]\begin{center}\scalebox{0.5}[0.5]{\rotatebox{270}{
\includegraphics{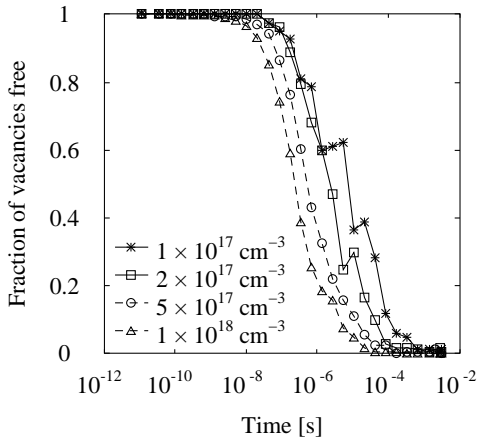}}}
\caption{\label{fig_V_freeV_C}
Fraction of vacancies free at various concentrations, with 18 shell interactions at 900 K.}
\end{center}\end{figure}
Figure \ref{fig_V_freeV_C} shows that, at higher concentrations, a greater fraction of 
vacancies is likely to be trapped in clusters.  Section \ref{sec_V_T} demonstrated 
that clusters have longer lifetimes, on average, at 900 K than at higher temperatures, 
so the growth of clusters at higher concentrations is most likely due to the aggregation 
of smaller 
clusters, while at lower concentrations, the number of clusters stays relatively constant 
while the 
mean size increases and the number of free vacancies decreases.  This implies that the 
capture of free vacancies drives cluster growth at lower concentrations. 
As with the temperature effects in section \ref{sec_V_T}, a cluster growth model which 
assumes either the aggregation or dissolution of small clusters across all concentrations 
will not accurately account for free vacancies in the system.

\subsection{Interaction range\label{sec_V_r}}
\begin{figure}\begin{center}\scalebox{0.5}[0.5]{\rotatebox{270}{
\includegraphics{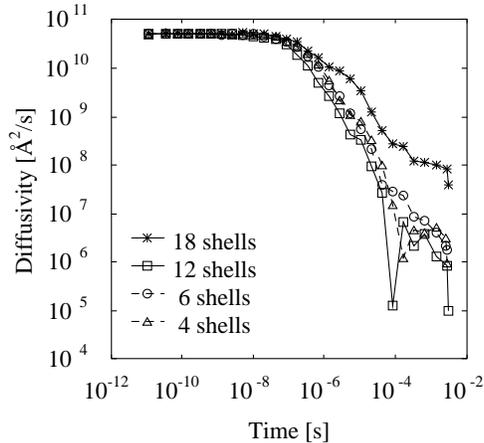}}}
\caption{\label{fig_V_DV_r}
Vacancy diffusivity at various interaction ranges, with $10^{18}\ \text{V cm}^{-3}$ at 
900 K.}
\end{center}\end{figure}
The simulations presented in this section were performed with a vacancy concentration of 
$10^{18}\ \text{cm}^{-3}$, in a box 102 unit cells in each dimension,  at 900 K.
The interaction range varied from four to eighteen lattice 
shells, using truncated and shifted forms of the vacancy pair interaction shown in 
Figure \ref{fig_potentials}.  Figure \ref{fig_V_DV_r} shows the vacancy diffusivity 
over time. 
\begin{figure}\begin{center}\scalebox{0.5}[0.5]{\rotatebox{270}{
\includegraphics{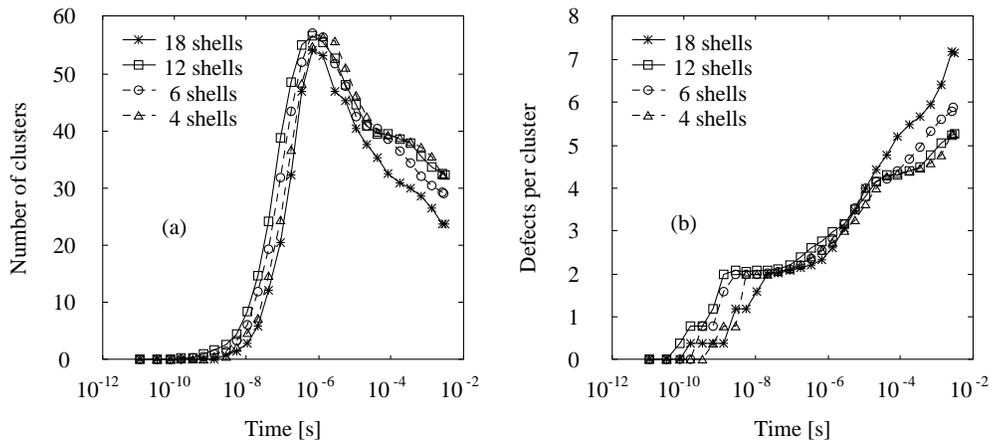}}}
\caption{\label{fig_V_nclust_ndefs_r}
Number of clusters (a) and cluster size (b) at various interaction ranges, 
with $10^{18}\ \text{V cm}^{-3}$ at 900 K.}
\end{center}\end{figure}
\begin{figure}\begin{center}\scalebox{0.5}[0.5]{\rotatebox{270}{
\includegraphics{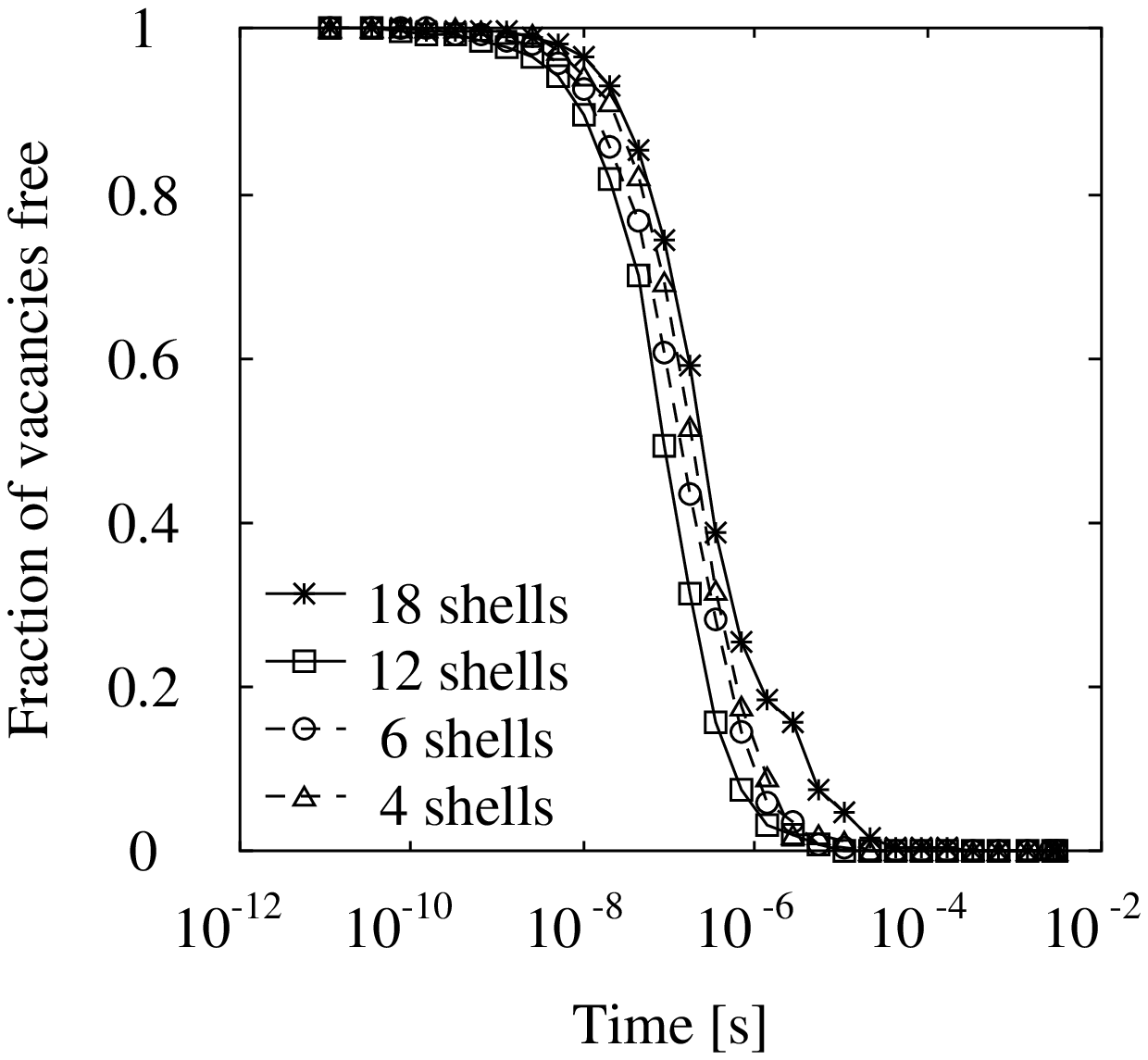}}}
\caption{\label{fig_V_freeV_r}
Fraction of vacancies free at various interaction ranges, with 
$10^{18}\ \text{V cm}^{-3}$ at 900 K.}
\end{center}\end{figure}
Initially vacancies diffuse freely for all interaction ranges considered.  As clusters 
form, the diffusivity decreases.  The decrease is not as pronounced for the longest 
interaction range considered, 18 shells, at which the power law exponent $\gamma = -0.8$, 
as at the shorter interaction ranges, where the exponent $\gamma = 1.15\pm0.15$.
This difference in $\gamma$ leads, at long times, to a diffusivity which is 
roughly two orders of magnitude greater, with 18 shell interactions, than the diffusivity
calculated with shorter range interactions.
The number of clusters and cluster size, shown in Figure \ref{fig_V_nclust_ndefs_r}a and b 
respectively, show a slight difference in long and short range interactions.  The
longest range interactions tend toward slightly larger, fewer clusters.  
Figure \ref{fig_V_freeV_r} shows the fraction of vacancies free, which is similar 
over time for the interaction ranges considered.  From these trends, a large interaction 
range leads to greater vacancy diffusivity over time, which produces fewer clusters.

\section{Conclusions}
We have presented results of kinetic lattice Monte Carlo simulations using long range
interactions, to the 18th nearest lattice neighbor, and large numbers of vacancies, 
simulated over relatively long time.  We studied vacancy clustering behavior as a function 
of temperature, concentration, and interaction range, visiting defects in random order at 
each simulation step.
Higher temperatures led to fewer, but larger, vacancy clusters, which existed for shorter 
times, on average, than clusters formed at lower temperatures and which grew by attracting 
free vacancies.  At lower temperatures, clusters grew by the dissociation and aggregation 
of smaller clusters.  This difference in growth mechanism has not been reported or
incorporated in any of the previously referenced models.
For higher vacancy concentrations, more clusters formed than at lower concentrations, with 
fewer free vacancies, but the average size of the clusters did not depend on concentration.
This result has not been previously reported.
A longer interaction range led to greater diffusivity with fewer clusters.
\par
We have also demonstrated that vacancy diffusivities are time dependent,
decreasing over time according, approximately, to a power law, $D(t) \sim t^{-\gamma}$.  
When the interaction range was long, 18 shells, $\gamma$ increased with increasing 
concentration, keeping temperature constant, and decreased with increasing temperature at 
fixed concentration.   Shorter interaction ranges led to greater values of $\gamma$.
The number and size of vacancy clusters was also shown to depend on time, in addition to
temperature and concentration.  None of the previously referenced models have explored
these time dependences.

\begin{acknowledgments}
We thank Art Voter and Blas Uberuaga for insightful discussion.  Initial parts of the work
by KMB were conducted in the Semiconductor Products sector of Motorola Inc.
The latter stages of
the work were conducted under LANL contract 25110-001-05, through the UC Davis-LANL
Materials Design Institute.  
\end{acknowledgments}


\end{document}